\documentclass[11pt]{article}

\usepackage{amsfonts}

\newcommand{\cM}{{\cal M}}

\begin{document}
\renewcommand{\thesection}{\Roman{section}}


\setcounter{page}{0}
\thispagestyle{empty}

ULM--TP/97--6 \hfill

July 1997

\normalsize

\vspace{0.5cm}

\renewcommand{\thefootnote}{\fnsymbol{footnote}}

\begin{center}  \huge \bf
On the Rate of Quantum Ergodicity on hyperbolic Surfaces and Billiards
\end{center}

\setcounter{footnote}{1}

\begin{center}
   by\\
   \vspace{4ex}

   {\large \bf R.\,Aurich%
\footnote{E-mail address: aurich@physik.uni-ulm.de}
       and M.\,Taglieber%
\footnote{E-mail address: tag@physik.uni-ulm.de}
          } \\
   \vspace{2ex}
   \small
   Abteilung Theoretische Physik, Universit\"at Ulm\\
   Albert-Einstein-Allee 11, D-89069 Ulm \\
   Federal Republic of Germany\\
\end{center}

\renewcommand{\thefootnote}{\arabic{footnote}}
\setcounter{footnote}{0}

\vspace{0.5cm}

\underline{\bf Abstract:}

The rate of quantum ergodicity is studied for three strongly chaotic
(Anosov) systems.
The quantal eigenfunctions on a compact Riemannian surface of genus $g=2$
and of two triangular billiards on a surface of constant negative curvature
are investigated.
One of the triangular billiards belongs to the class of arithmetic systems.
There are no peculiarities observed in the arithmetic system
concerning the rate of quantum ergodicity.
This contrasts to the peculiar behaviour with respect to the statistical
properties of the quantal levels.
It is demonstrated that the rate of quantum ergodicity in the three
considered systems fits well with the known upper and lower bounds.
Furthermore, Sarnak's conjecture about quantum unique ergodicity
for hyperbolic surfaces is confirmed numerically in these three systems.

\vspace{1cm}
\fbox{\begin{minipage}{15cm}
{\Large \bf \underline{Note:}} \\

{\large \centerline{The postscript file of this paper containing all figures
is available at: }}

\vspace*{3pt}
{\Large \centerline{http://www.physik.uni-ulm.de/theo/qc/}}

\vspace*{8pt}
\end{minipage}}


\newpage

\section{Introduction}

In classical mechanics the behaviour of systems can be classified
in the range from integrable towards strongly chaotic.
Thereby one of the lowest suppositions of chaos is given by ergodicity
being followed by the mixing property and finally by the Anosov property.
Thus as a first step towards chaos it suggests itself to study the
quantum mechanical analogue of the classical ergodicity
theorem (see e.g.\ \cite{Pollicott93}).

Here, we restrict us to bounded Hamiltonian systems having a purely
discrete quantal eigenvalue spectrum $\{ E_n \}$, $n \in \mathbb{N}$,
with corresponding eigenstates $\Psi_n$ which are assumed to be
orthonormalized with respect to the L$_2$ norm.
Especially, we are concerned with the eigenstates of the
Laplace-Beltrami operator $\Delta$ on compact Riemannian manifolds $\cM$
with constant negative curvature, i.\,e.\
\begin{equation}
- \, \Delta \Psi_n \; = \; E_n \, \Psi_n
\hspace{10pt} .
\end{equation}
In quantum mechanics the following measure was introduced for compact
Riemannian manifolds $\cM$
\cite{Zelditch94a,Zelditch94b}
\begin{equation}
\label{Def_Sk}
S_k(E;A) \; := \; \frac{1}{N(E)} \; \sum_{E_n\le E}
\; \left| \; \left( \, A \Psi_n,\Psi_n \right)\, - \overline{\sigma}_A
\; \right|^k
\hspace{10pt} , \hspace{10pt} k > 0
\hspace{10pt} ,
\end{equation}
where $N(E) := \#\{E_n | E_n\leq E\}$ is the staircase function counting
the number of the quantal levels $E_n$ below the energy $E$.
The operator $A$ is a pseudo-differential operator of $0$-th order,
i.\,e.\ roughly speaking, it is an operator containing no derivatives.
For an introduction to the theory of pseudo-differential operators,
see \cite{Hoermander,Folland}.
The average $\overline{\sigma}_A$ is defined as
\begin{equation}
\overline{\sigma}_A := \frac{1}{\hbox{vol}(S^\star \cM)} \;
\int_{S^\star \cM} \; \sigma_A \; d\tau
\hspace{10pt} ,
\end{equation}
where $S^\star \cM$ is the unit cotangent bundle,
i.\,e.\ the energy shell in phase space.
The Liouville measure on $S^\star \cM$ is denoted by $d\tau$,
and $\sigma_A$ is the principal symbol of $A$, which can be considered
as the classical analogue of $A$.
If the flow on $S^\star \cM$ is ergodic
it is proven \cite{Schnirelman74,Zelditch87,CdVeriere85,ZelZwo96} that
\begin{equation}
\label{Quantum_Ergodicity}
S_k(E;A) \; \to \; 0
\hspace{10pt} \hbox{ for } \hspace{10pt}
E \rightarrow \infty \hspace{10pt} .
\end{equation}
Thus the vanishing of (\ref{Def_Sk}) in the limit $E \rightarrow \infty$
can be considered as a quantum analogue of ergodicity.

The definition of {\it quantum ergodicity} as (\ref{Quantum_Ergodicity})
is equivalent to another formulation.
For bounded sequences $\{a_n\}$, $a_n\in \mathbb{C}$, one has the
following lemma \cite{Walters82}.
The condition
$$
\lim_{N\to\infty} \; \frac 1 N \; \sum_{n\leq N} \; | \, a_n \, |^p \;
= \; 0
\hspace{10pt} \hbox{ for any } \; \; p > 0
$$
is equivalent to
$$
\lim_{j\to\infty} a_{n_j} \; = \; 0 \hspace{10pt} ,
$$
where $\{n_j\}$ is a subsequence of $\mathbb{N}$ of density 1.
A subsequence $n_j$, $n_1<n_2<\dots$, is defined to be of density $d$
$(0\leq d \leq 1)$ if
$$
\lim_{N\to\infty} \; \frac {\#\{n_j | n_j \leq N\}}
{\#\{n | n \leq N\}} \; = \; d \hspace{10pt} .
$$
Thus the condition (\ref{Quantum_Ergodicity}) of quantum ergodicity reads as
\begin{equation}
\label{Limit}
\left( \, A \Psi_{n_j},\Psi_{n_j} \right) \; \rightarrow \;
\overline{\sigma}_A \hspace{10pt} \hbox{ for } \hspace{10pt}
E_{n_j}\; \to \; \infty
\end{equation}
for a subsequence $\{n_j\}$ of density one.
The kind of the decay of (\ref{Def_Sk}) for $E \rightarrow \infty$
and (\ref{Limit}) is coined as the {\it rate} of quantum ergodicity.
Quantum {\it unique} ergodicity happens to be if the condition
(\ref{Limit}) holds for the complete sequence $\{ n\}$
\begin{equation}
\label{Unique}
\left( \, A \Psi_n,\Psi_n \right) \; \rightarrow \;
\overline{\sigma}_A \hspace{10pt} \hbox{ for } \hspace{10pt}
E_n \; \to \; \infty
\hspace{10pt} .
\end{equation}

Attention has to be paid for interpreting the condition
(\ref{Quantum_Ergodicity}) as a measure for quantum ergodicity,
since one cannot conclude from
(\ref{Quantum_Ergodicity}) that the classical system is ergodic.
An instructive example is the sphere $S^2$, which can be
considered as the counterpart of the surfaces studied in this paper
since the sphere $S^2$ has constant positive curvature.
The usual basis of spherical harmonics $\{Y_l^m\}$ is indeed not
quantum ergodic as becomes clear by considering subsequences
of a fixed rational $m/l$.
However, for almost all orthogonal bases the eigenfunctions of the sphere
$S^2$ are quantum ergodic \cite{Zelditch92}.
An example of a subsequence of density zero violating (\ref{Limit}) is provided
by the subsequence $\{Y_l^l\}$ which is very strongly localized at the equator.
If one considers, however, the subsequence $\{Y_l^m\}$ with
$m/l \geq c$, $0<c<1$, one obtains a subsequence of density
$\frac{1-c}{2} > 0$ which is concentrated in the neighbourhood
of the equator given by $\sin\theta \geq c$ \cite{CdVeriere89}.

In the case of compact Riemannian manifolds of negative curvature
the best known upper bound of $S_k(E;A)$ is given by \cite{Zelditch94a}
\begin{equation}
\label{Sk_upper_bound}
S_k(E;A) \; = \; O\left( (\log E)^{-\frac k 2}\, \right)
\hspace{10pt} .
\end{equation}
By considering the sum over eigenstates in eq.(\ref{Def_Sk})
one avoids possible sparse
subsequences $\{n_j\}$ of zero density, for which
$$
\left| \;  \left( \, A \Psi_{n_j},\Psi_{n_j} \right)\, - \overline{\sigma}_A
\; \right| \; = \; \Omega(1)
\hspace{10pt} ,
$$
which might correspond to eigenstates scarred by short classical
periodic orbits.
However, this does not occur for the hyperbolic octagons studied in
\cite{Aurich95a}.
Here $\Omega$ is the Hardy-Littlewood-lower-bound symbol
being defined as
$$
f(x) = \Omega(g(x)) \hspace{10pt} \hbox{ for } \hspace{10pt}
x \rightarrow \infty \hspace{10pt} \hbox{ if } \hspace{10pt}
f(x) \neq o(g(x))
\hspace{10pt} .
$$
If the Liouville average of the sub-principal symbol
$\sigma_{\hbox{\scriptsize sub}}(A\,)$ does not vanish,
then a lower bound is given by \cite{Zelditch94b}
\begin{equation}
\label{lower_bound}
\left| \;  \left( \, A \Psi_n,\Psi_n \right)\, - \overline{\sigma}_A
\; \right| \; = \; \Omega\left( E_n^{-\frac 1 2}\, \right)
\hspace{10pt} .
\end{equation}
Addressing the question of quantum ergodicity in individual eigenstates,
a conjecture is put forward by Sarnak \cite{Sarnak95} which states
that for hyperbolic surfaces in which the curvature is negative
\begin{equation}
\label{Sarnak_Conjecture}
\left( \, A \Psi_n,\Psi_n \right)\, - \overline{\sigma}_A
\; = \; O\left( E_n^{-\frac 1 4 + \varepsilon}\, \right)
\hspace{10pt} \hbox{ for all } \hspace{10pt} \varepsilon > 0
\hspace{10pt} .
\end{equation}

The main topic in this paper deals with the rate of the decay of $S_1(E;A)$.
An extended analysis including higher $S_k(E;A)$, $k\geq 1$, and other
statistics of the wave functions of the hyperbolic billiards considered in this
paper
can be found in \cite{Taglieber97}.
A detailed analysis concerning the behaviour of $S_1(E;A)$ of Euclidean
billiards,
i.\,e.\ the stadium billiard, the cardioid billiard and the so-called cosine
billiard,
can be found in \cite{Baecker97}.

\section{The chaotic systems}

\label{Modell_Systems}

Surfaces of constant negative Gaussian curvature constitute
a simple model for chaos
since the classical dynamics on these surfaces possess all of the
characteristic properties of instability:
ergodicity, mixing and the Anosov property.
As a model for this surface we use the Poincar\'e disc ${\cal D}$
which consists of the interior of the unit circle in the
complex $z$--plane $(z=x_1+i x_2)$ endowed with the hyperbolic metric
\begin{equation}
\label{Metric}
g_{ij} \; = \; \frac{4}{(1-x_1^2-x_2^2)^2} \delta_{ij} \hspace{15pt}
, \hspace{15pt} i,j = 1,2
\end{equation}
corresponding to constant negative Gaussian curvature $K=-1$.
This fixes the length scale.
The systems which are studied in this paper are defined on this hyperbolic
surface.

The classical motion (geodesic flow) is determined
by the Hamiltonian $H=\frac{1}{2m} p_i g^{ij} p_j$,
$p_i \, = \, m \, g_{ij} \, dx^j/dt$.
The geodesics are circles intersecting the boundary
of the Poincar\'e disc ${\cal D}$ perpendicularly.

The quantum mechanical system is governed
by the Schr\"odinger equation
\begin{equation}
\label{Schrod}
-\Delta \Psi_n(z) = E_n \Psi_n(z)
\hspace{15pt} , \hspace{15pt}
\Delta = \frac 1 4 (1-x_1^2-x_2^2)^2 \,
\left( \frac{\partial^2}{\partial x_1^2} +
       \frac{\partial^2}{\partial x_2^2} \right)
\hspace{10pt} ,
\end{equation}
where $\Delta$ denotes the non--Euclidean Laplacian corresponding to
the hyperbolic metric (\ref{Metric}).
Here and in the following units $\hbar=2m=1$ are used.
The wave functions $\Psi_n(z)$ have to obey appropriate boundary conditions,
see below.

Since we are interested in possible differences in the rate of
quantum ergodicity between non-arithmetic and arithmetic systems,
we choose the asymmetric octagon shown in figure \ref{kreis_m31_g01}
as the non-arithmetic system,
and the so-called regular octagon for the arithmetic system.
Both are Riemannian surfaces of genus two.
(For more details about these Riemannian surfaces, see e.\,g.\
\cite{Balazs86,Gutzwiller90}.)
However, both systems decompose into irreducible symmetry representations
such that we consider only one symmetry class.
In the case of the non-arithmetic octagon which only possesses one involution
symmetry the positive symmetry class is chosen, i.\,e.\ $\Psi(-z)=+\Psi(z)$.
The regular octagon
which has the highest symmetry possible for a surface of genus 2,
decomposes into 96 symmetry classes from which the one is selected
which corresponds to a hyperbolic triangle with angles
$\alpha=\pi/3$, $\beta=\pi/8$ and $\gamma=\pi/2$.
The sides $a,b$ and $c$ being opposite to their corresponding angles
are endowed with the boundary conditions Dirichlet, Neumann and Neumann,
respectively.
Since these two desymmetrized systems differ in some aspects,
we incorporate in our study a further triangular billiard
which differs only in the imposed boundary conditions from the former.
For the sides $a,b$ and $c$ the boundary conditions Neumann, Neumann
and Dirichlet are chosen, respectively,
which do not provide a subspectrum of the regular octagon.
The main point is that this combination of boundary conditions
leads to a non-arithmetic system.
For a detailed investigation of these two hyperbolic triangular billiards
with respect to the quantal level statistics, see \cite{Aurich95}.

Thus our study is based on one parity class of a non-arithmetic
Riemannian surface and on one hyperbolic triangle endowed with two
different boundary conditions from which only one is arithmetic.
In the following the non-arithmetic billiard is called billiard ${\cal A}$
and the other billiard ${\cal B}$, see figure \ref{kreis_m31_g01}.
Using the boundary element method \cite{Aurich93}
the first 3000 wave functions of the positive symmetry class of
the asymmetric octagon are computed.
For billiard ${\cal A}$ and ${\cal B}$ the first 2092 and 2099 wave functions
are determined, respectively.
The following statistical analysis is based on this large set of
wave functions.

\section{On the behaviour of $S_1(E;A)$ in the configuration space}

\label{behaviour_S_1}
The pseudo-differential operator of 0-th order $A$ is chosen to
be the multiplication operator with
the characteristic function of a subset ${\cal X} \subset {\cal M}$
on the Poincar\'e disc ${\cal D}$, i.\,e.\ for $z \in {\cal X}$
it is one and zero otherwise.
Thus $A$ does not depend on the momentum $p$.
Then the scalar product in (\ref{Def_Sk}) reduces to
\begin{equation}
\label{DefScalar}
(\, A \Psi_n, \Psi_n ) \; = \;
\int_{\cal X} \, d\mu(z) \, | \Psi_n(z) |^2
\hspace{10pt} ,
\end{equation}
i.\,e.\ the probability to find the particle in ${\cal X}$.
The invariant volume element on ${\cal D}$ is designated as $d\mu(z)$.
With this choice of $A$ the average $\overline{\sigma}_A$ is given by
$$
\overline{\sigma}_A = \frac{\hbox{Area}({\cal X})}{\hbox{Area}({\cal M})}
\hspace{10pt} ,
$$
where $\hbox{Area}({\cal X})$ is the area of the domain ${\cal X}$ and
$\hbox{Area}({\cal M})$ is the area of the desymmetrized domain.
The average $\overline{\sigma}_A$ is exactly what is classically expected
for the mean value of the classical analogue of the observable $A$
for an ergodic system.
The domain ${\cal X}$ is chosen to be a circular domain.
For the numerical integration we choose a disc with radius $R$
having its center at $z=0$, i.\,e.\ at the origin in the Poincar\'e disc.
The disc is then transformed by a linear fractional transformation
$z \to (\alpha z + \beta) / (\beta^\star z + \alpha^\star)$,
$\alpha,\beta \in \mathbb{C}$.
This maps the circle centered at $z=0$ conformally to the domain ${\cal X}$
inside
the billiards for suitably chosen $\alpha$ and $\beta$.
An example of such a domain ${\cal X}$ characterizing the
operator $A$ is shown in figure \ref{kreis_m31_g01} for the asymmetric octagon
as well as for the triangular billiards.
These two operators are chosen for the examples shown in
figures \ref{an_sarnak_conjecture}, \ref{cn_m31_g01}, \ref{Distr_cn_m31_g01},
\ref{S1_Oktagon_G01} and \ref{Pic_an_orbit}.

In order to understand $S_1(E;A)$ let us consider the summands in
(\ref{Def_Sk}) in more detail and define
\begin{equation}
\label{Def_an}
\tilde a_n \; := \; \, (\, A \Psi_n, \Psi_n ) - \overline{\sigma}_A
\hspace{10pt} \hbox{ and } \hspace{10pt}
a_n \; := \; | \, \tilde a_n \, |
\hspace{10pt} .
\end{equation}
Dealing numerically with finite energies we consider the limit
$\varepsilon\to 0$ in Sarnak's conjecture (\ref{Sarnak_Conjecture})
and introduce
\begin{equation}
\label{Def_cn}
\tilde c_n \; := \; \, \tilde a_n \, E_n^{1/4}
\hspace{10pt} \hbox{ and } \hspace{10pt}
c_n \; := \; | \, \tilde c_n \, |
\hspace{10pt} .
\end{equation}
Thus the mean behaviour of $a_n$ is given by
$\overline{c}\, E_n^{-1/4}$ with $\overline{c}$ being the mean value of $c_n$.
For the domains ${\cal X}$ shown in figure \ref{kreis_m31_g01} we average the
$a_n$'s with a triangular shaped window function,
denoted by $\left< \dots \right>$, with a base width of $\Delta n=100$
and fitted them to $f(E) = c_f E^{-1/4}$ with $c_f$ being a fit parameter.
The result is shown in figure \ref{an_sarnak_conjecture} for all three systems.
One observes a good agreement in favor of Sarnak's conjecture and notes
that the lower bound (\ref{lower_bound}) is safely fulfilled.

The first 3000 numbers $\tilde c_n$, which are expected to be
pseudo-random numbers with fluctuations independent of $n$,
are shown in figure \ref{cn_m31_g01} in the case of the asymmetric octagon.
One observes that the amplitude is of the same order over the whole range.
Furthermore, no large exceptions are observed in agreement with Sarnak's
conjecture.
Also for billiard ${\cal A}$ and ${\cal B}$ no such exceptions are discovered.

Numerically one observes a Gaussian distribution of the $\tilde c_n$'s.
For finite energies this distribution has, however, a non-zero mean.
This is due to the imposed boundary conditions for the desymmetrized systems.
In the desymmetrized asymmetric octagon the elliptic points,
denoted in figure \ref{kreis_m31_g01} as dots, play a special role.
Because the hyperbolic octagons are Riemannian surfaces without boundary,
there are no Neumann or Dirichlet boundary conditions
which influence the eigenstates as in the case of the triangular
billiards.
It turns out that for the negative parity class, the eigenstates must
vanish at the elliptic points, whereas in the case of positive parity,
the eigenstates possess local extrema there.
In the case of the regular octagon we consider as a special desymmetrized
class the triangular billiard discussed in section \ref{Modell_Systems}.
Here Dirichlet and Neumann boundary conditions influence the eigenstates
significantly near the boundaries.
The special role of the boundary influences can be very clearly seen by
considering the sum over the modulus of the eigenstates
\begin{equation}
\label{WFK_Sum}
W(z,E) \; := \; \sum_{E_n\leq E} \; |\, \Psi_n(z)\, |^2
\hspace{10pt} .
\end{equation}
In figure \ref{Pic_WFK_Sum} this quantity is shown for the positive parity
class of the asymmetric octagon, the triangular billiard ${\cal A}$ with
boundary
conditions incompatible with the regular octagon and for the
triangular billiard ${\cal B}$ derived from the regular octagon.
One observes deviations from an equidistribution enforced by the
imposed boundary conditions near the elliptic points and near the boundary,
respectively.
At the Neumann boundary the amplitude is twice the mean value
whereas at the Dirichlet boundary the amplitude vanishes, of course.
For more details, see \cite{Hoermander}.
The crucial point is that these boundary effects slightly modify the
local expectation value of $|\Psi_n(z)|^2$ far away from the boundary.
This is the reason for the observed non-zero mean in the distribution
of the $\tilde c_n$.
Alternatively, this shift can also be taken into account by using orbit
corrections
to $\overline{\sigma}_A$ in equation (\ref{Def_Sk})
as discussed in section \ref{Orbit_Theory}.
With increasing energy these modification decrease
but since one is numerically restricted to finite energies,
we take the shift in the Gaussian into account.
For the asymmetric octagon the cumulative distribution of $\tilde c_n$
is shown in figure \ref{Distr_cn_m31_g01} in comparison with
\begin{equation}
\label{fit_erfc}
I_f(\tilde c\,) = \frac 1 2 \hbox{erfc}( -(\tilde c\,-\delta) /
\sqrt{2} \beta )
\end{equation}
where $\beta$ is the standard deviation and $\delta$ the mean value
of the distribution of $\tilde c_n$.
A fit excluding the first 50 $\tilde c_n$'s
yields the values $\beta = 0.0957\dots$ and $\delta =-0.0029\dots$
which is very small in comparison with $\beta$.
A Kolmogorov-Smirnov test gives a significance level of $55\%$.
Neglecting the shift $\delta$ leads to a significance level
being almost zero.
In the case of billiard ${\cal A}$ one obtains for the domain shown in figure
\ref{kreis_m31_g01} $\beta = 0.5154\dots$ and $\delta =-0.0277\dots$
again with a significance level of 55\%.
Billiard ${\cal B}$ gives $\beta = 0.4333\dots$ and $\delta =-0.0237\dots$
with a significance level of 94\%.
The Gaussian behaviour of $\tilde c_n$ implies for $c_n$
the distribution density
\begin{equation}
\label{Gauss_half}
P(c) \; = \; \frac{1}{\sqrt{2 \pi} \beta} \; \left\{
e^{-\frac{(c-\delta)^2}{2\beta^2}} \; + \;
e^{-\frac{(c+\delta)^2}{2\beta^2}} \; \right\}
\hspace{10pt} \hbox{ for } \hspace{10pt} c \; \geq 0
\hspace{10pt} .
\end{equation}
Thus $\beta$ and $\delta$ determine the mean value $\overline{c}$, i.\,e.\
\begin{equation}
\overline{c} \; = \; \sqrt{\frac{2}{\pi}} \; \beta \;
e^{-\frac{\delta^2}{2\beta^2}} \; + \;
\delta \; \hbox{erf}\left(\frac{\delta}{\sqrt 2 \beta}\right)
\; \simeq \; \sqrt{\frac{2}{\pi}} \; \beta
\hspace{10pt} \hbox{ for } \hspace{10pt} |\,\delta\,| \ll \beta
\hspace{10pt} .
\end{equation}
Furthermore, let us define
\begin{equation}
\label{Def_zn}
z_n \; := \; a_n - \overline{c} \, E_n^{-\frac 1 4}
\end{equation}
which can be considered as a random variable with zero mean.
The standard deviation $\gamma$ of the distribution of
$z_n E_n^{1/4} = c_n - \overline{c}$ is then given by
\begin{equation}
\gamma \; = \;
\sqrt{ \beta^2 + \delta^2 - \overline{c}^2 }
\; \simeq \;  \sqrt{1-\frac 2 \pi} \; \beta
\hspace{10pt} \hbox{ for } \hspace{10pt} |\,\delta\,| \ll \beta
\hspace{10pt} .
\end{equation}
With the definitions $\overline{a}_n := \overline{c} \, E_n^{-\frac 1 4}$
and (\ref{Def_zn}), one has, replacing $N(E)$ in (\ref{Def_Sk}) by
$\overline{N}(E)$,
\begin{equation}
\label{S1_split}
S_1(E;A) \; \simeq \; \frac{1}{\overline{N}(E)} \;
\left( \, \sum_{E_n<E} \; \overline{a}_n \; + \;
\sum_{E_n<E} z_n \right)
\hspace{10pt} .
\end{equation}
The first sum describes the mean behaviour $\overline{S}_1(E;A)$
whereas the second sum fluctuates around zero.

The mean behaviour $\overline{S}_k(E;A)$ is determined by Weyl's law
\begin{equation}
\label{Weyls_law}
\overline{N}(E) \; \simeq \; \frac{\hbox{Area}({\cal M})}{4\pi} \; E
\hspace{10pt} , \hspace{10pt} E \; \to \infty \hspace{10pt} ,
\end{equation}
neglecting the circumference term in the case of billiards ${\cal A}$
and ${\cal B}$, leading to
\begin{equation}
\label{S1_mean}
\overline{S}_1(E;A) \; = \;
\frac{1}{\overline{N}(E)} \, \sum_{E_n<E} \overline{a}_n
\; \simeq \; \frac{4\, \overline{c}}{3} \; E^{-\frac 1 4}
\hspace{10pt} .
\end{equation}

Now let us consider the deviations from this mean behaviour described
by the second sum in (\ref{S1_split}).
Assuming that the $z_n$'s are independent random variables,
the variance $V(E)$ of the sum $\sum_{E_n<E} z_n$
is given by $V(E) = \sum_{E_n<E} \nu_n^2$
with $\nu_n = \gamma \, E_n^{-1/4}$.
The variance $V(E)$ can be estimated using Weyl's law (\ref{Weyls_law})
\begin{equation}
\label{zn_variance}
V(E) \; = \; \sum_{E_n<E} \, \gamma^2 E_n^{-\frac 1 2}
\; \simeq \; \frac{2 \overline{N}(E) \gamma^2}{\sqrt{E}}
\hspace{10pt} .
\end{equation}
Thus $S_1(E;A)$ computed directly from the wave functions should
lie within the range
\begin{equation}
\label{S1_model}
S_1^{\hbox{\scriptsize model}}(E;A) \; := \;
\frac{4\, \overline{c}}{3} \; E^{-\frac 1 4} \; \pm \;
\frac{\sqrt 2 \gamma}{\sqrt{\overline{N}(E)} \, E^{\frac 1 4}}
\hspace{10pt} ,
\end{equation}
if the deviations are compared with one standard deviation and
if the above assumptions are correct.
This curve is determined from the two parameter $\overline{c}$ and $\gamma$
which in turn are determined by $\beta$ and $\delta$.
A comparison of (\ref{S1_model}) with the directly computed $S_1(E;A)$
provides in this way also a test of the exponent $\frac 1 4$
in (\ref{Sarnak_Conjecture}).
Figure \ref{S1_Oktagon_G01} demonstrates in the case of the asymmetric octagon
that the rate of quantum ergodicity for $S_1(E;A)$
is indeed given by the power $\frac 1 4$.
In contrast to the discussion above, the shown model curve
contains the first $k$ true $a_n$'s, i.\,e.\
\begin{eqnarray}
\nonumber
\tilde S_1^{\hbox{\scriptsize model}}(E;A) & = &
\frac{1}{\overline{N}(E)} \; \left( \, \sum_{E_n\leq E_k} \; a_n \; + \;
\sum_{E_k<E_n\leq E} \overline{a}_n \right) \\
\label{S1_prakt}
& = &
\frac{1}{\overline{N}(E)} \; \sum_{E_n\leq E_k} \; a_n \; + \;
\frac{4 \overline{c}}{3} \, \left[ E^{-1/4} - E_{k}^{-1/4} \right]
\hspace{10pt} .
\end{eqnarray}
This modification was necessary since
for a given domain ${\cal X}$ the lowest wave functions have not enough
structure
within ${\cal X}$ in order to provide pseudo-random numbers $z_n$.
The value of $k$ depends on the size of the domain ${\cal X}$.
In this case $k=50$ was chosen but already $k=10$ gives
a reasonable approximation.
Note that 3000 wave functions are included.

Alternatively, we study the rate of decline of $S_1(E;A)$
by fitting to $S_1(E;A)$ the function
\begin{equation}
\label{Fit}
f(E) \; := \; \alpha \, E^{-\rho}
\end{equation}
and compare $\rho$ with the expected $\frac 1 4$.
For our three systems we choose a sequence of ten domains ${\cal X}_i$.
Ten circles centered at the origin of the Poincar\'e disc are chosen
which are mapped by the same linear fractional transformation
such that the considered domains do not contain elliptic points.
Within a sequence the radius of the domains are chosen such that the area of
the
domains increases in constant steps of $1/10$-th of the area of the largest
domain.
The two sequences are shown in figure \ref{Pic_circle_series} for
the asymmetric octagon and for the triangular billiards.
In the case of the asymmetric octagon the selected domains ${\cal X}_i$ have
no points which are mapped by the transformation $z\to -z$
to points within the same domain
thus eliminating dependencies by the involution symmetry.
In figure \ref{Pic_S1_series} the obtained $S_1(E;A)$ are shown
together with the fit (\ref{Fit}).
The functions (\ref{Fit}) with the fitted parameters are only shown
over the interval used for the fit.
The values of the fit parameter $\alpha$ and $\rho$ are given in
table \ref{Table_fit_parameter}.
One observes for the values of $\rho$ a good agreement with the conjectured
value $\frac 1 4$.
This result clearly shows that the upper bound (\ref{Sk_upper_bound}) is
valid for the energies considered here.
The agreement is better in the case of the asymmetric octagon.
In the case of the two triangular billiards slightly larger deviations occur.
In table \ref{Table_fit_parameter_nvar} the values of the fit parameter
$\rho$ are shown depending on the interval for which the fit is carried out.
A tendency is observed that $\rho$ approaches the conjectured value
$\frac 1 4$ for higher energies.
Using domains ${\cal X}_i$ which are centered at other locations in ${\cal M}$
leads to other values of $\rho$, some above $\frac 1 4$ and some below.
It turns out that the average over these $\rho$'s is very close to
$\frac 1 4$ as it is demonstrated for ten domains having different centers
in \cite{Taglieber97}.
Thus numerical evidence is given in favor of the conjecture.

The value of $\alpha$ depends on the area as shown in figure \ref{Pic_alpha}.
In order to compare the three systems we plotted $\alpha$ versus the
index $i=1,\dots,10$ of the domains ${\cal X}_i$.
Since Area(${\cal X}_i$) increases linearly with $i$, $\alpha$ is
also plotted versus the area.
One observes a linear behaviour of $\alpha$ over a wide range.
The asymmetric octagon shows the linear behaviour up to $i=7$,
and the curve belonging to billiard ${\cal B}$ flattens at $i\simeq9$.
In the limit Area$({\cal X}) \to$ Area$({\cal M})$ the values of $a_n$
decrease rapidly because of the normalization of the
wave functions with respect to ${\cal M}$.
This in turn leads to a decline of $\alpha$ for areas ${\cal X}$ approaching
${\cal M}$.
Thus one has three different regions for $\alpha$, one linear region for small
domains, a transition region and finally a region with decreasing $\alpha$.

\begin{table}[htb]\center
\begin{tabular}{|c|c|c|}
\hline
asymmetric octagon & billiard ${\cal A}$ &  billiard ${\cal B}$ \\
\hline
\begin{tabular}{|c|c|c|}
Area & $\alpha$ & $\rho$ \\
\hline
0.2691  &   0.036  &   0.251  \\
0.5383  &   0.057  &   0.251  \\
0.8074  &   0.069  &   0.249  \\
1.0766  &   0.082  &   0.250  \\
1.3457  &   0.096  &   0.257  \\
1.6149  &   0.108  &   0.262  \\
1.8840  &   0.121  &   0.268  \\
2.1532  &   0.119  &   0.261  \\
2.4223  &   0.115  &   0.254  \\
2.6915  &   0.115  &   0.253  \\
\end{tabular} &
\begin{tabular}{|c|c|c|}
Area & $\alpha$ & $\rho$ \\
\hline
0.0031  &   0.101  &   0.234  \\
0.0063  &   0.122  &   0.209  \\
0.0094  &   0.168  &   0.215  \\
0.0126  &   0.223  &   0.224  \\
0.0157  &   0.266  &   0.226  \\
0.0189  &   0.309  &   0.229  \\
0.0220  &   0.360  &   0.234  \\
0.0252  &   0.419  &   0.239  \\
0.0283  &   0.480  &   0.244  \\
0.0315  &   0.550  &   0.249  \\
\end{tabular} &
\begin{tabular}{|c|c|c|}
Area & $\alpha$ & $\rho$ \\
\hline
0.0031  &   0.076  &   0.211  \\
0.0063  &   0.134  &   0.221  \\
0.0094  &   0.151  &   0.212  \\
0.0126  &   0.182  &   0.215  \\
0.0157  &   0.219  &   0.220  \\
0.0189  &   0.272  &   0.229  \\
0.0220  &   0.304  &   0.232  \\
0.0252  &   0.339  &   0.235  \\
0.0283  &   0.353  &   0.234  \\
0.0315  &   0.367  &   0.233  \\
\end{tabular} \\
\hline
\end{tabular}
\caption{\label{Table_fit_parameter}
The fit parameters $\alpha$ and $\rho$ are shown for the three systems
for the sequences of domains ${\cal X}_i$ shown in figure
\ref{Pic_circle_series}.
}
\end{table}

\begin{table}[htb]\center
\begin{tabular}{|c|c|}
\hline
billiard ${\cal A}$ &  billiard ${\cal B}$ \\
\hline
\begin{tabular}{|c|c|c|c|}
Area & $\rho_1$ & $\rho_2$ & $\rho_3$   \\
\hline
0.0031  &   0.218  &   0.236  &   0.234 \\
0.0063  &   0.177  &   0.206  &   0.235 \\
0.0094  &   0.163  &   0.220  &   0.243 \\
0.0126  &   0.160  &   0.229  &   0.259 \\
0.0157  &   0.154  &   0.230  &   0.273 \\
0.0189  &   0.150  &   0.237  &   0.267 \\
0.0220  &   0.141  &   0.244  &   0.255 \\
0.0252  &   0.146  &   0.252  &   0.254 \\
0.0283  &   0.151  &   0.258  &   0.254 \\
0.0315  &   0.150  &   0.259  &   0.260 \\
\end{tabular} &
\begin{tabular}{|c|c|c|c|}
Area & $\rho_1$ & $\rho_2$ & $\rho_3$   \\
\hline
0.0031  &   0.202  &   0.222  &   0.225 \\
0.0063  &   0.234  &   0.229  &   0.235 \\
0.0094  &   0.237  &   0.218  &   0.221 \\
0.0126  &   0.233  &   0.227  &   0.219 \\
0.0157  &   0.229  &   0.229  &   0.227 \\
0.0189  &   0.230  &   0.223  &   0.245 \\
0.0220  &   0.240  &   0.217  &   0.251 \\
0.0252  &   0.250  &   0.217  &   0.255 \\
0.0283  &   0.264  &   0.213  &   0.249 \\
0.0315  &   0.276  &   0.211  &   0.244 \\
\end{tabular} \\
\hline
\end{tabular}
\caption{\label{Table_fit_parameter_nvar}
The fit parameter $\rho$ is shown for the two billiards in dependence on
the fit interval: $\rho_1$ is fitted over $n=200\dots 800$,
$\rho_2$ over $n=800\dots 1400$ and $\rho_3$ over $n=1400\dots 2000$.
}
\end{table}

\section{The orbit theory}

\label{Orbit_Theory}

In this section we consider the corrections to $\overline{\sigma}_A$
which are provided by the orbit theory.
Furthermore, the question is posed for possible differences in the
properties of the wave functions between arithmetic and non-arithmetic systems.
This question is enforced by the fact that the properties of the energy level
statistics are quite different \cite{BogGeoGiaSch92,BolSteSte92}.
This fingerprint is caused by the exponentially degenerated length
spectrum of the periodic orbits in the case of arithmetic systems.
It is thus important to look at the description of the wave functions
in terms of classical orbits.

The orbit theory derived in \cite{Aurich91} allows to express a nearly
arbitrarily
weighted sum over wave functions by a corresponding sum over classical orbits.
Here, in contrast to the periodic orbit theory regarding the eigenvalue
spectrum, all non-periodic orbits are necessary, since it is not a
trace formula.
For compact Riemannian surfaces the orbit formula can be stated as
\begin{equation}
\label{Wave_Sum}
\sum_{n=0}^\infty \; h(p_n) \; \Psi^\star_n(z) \; \Psi_n(z') \; = \;
\frac{1}{2\pi} \; \sum_{b\in \Gamma} \;
\widehat{h}\left(\cosh d(z,b(z'))\right)
\hspace{10pt} ,
\end{equation}
where $h(p)$ is an even function, which is holomorphic in the strip
$|\Im p | \leq \frac 1 2 +\varepsilon, \varepsilon>0$
and of order $h(p) = O(p^{-2-\delta}), \delta>0$ for $|p|\to\infty$.
The sum on the right-hand-side runs over the Fuchsian group $\Gamma$
which defines the tessellation of the Poincar\'e disc ${\cal D}$.
Furthermore, $\widehat{h}(y)$ is the Mehler transform of $h(p)$
\begin{equation}
\widehat{h}(y) \; = \; \int_0^\infty \, dp \, p \, \tanh(\pi p) \,
P_{-\frac 1 2 + i p}(y) \, h(p)
\end{equation}
with $P_{-\frac 1 2 + i p}(y), y>1$, being the Legendre function
of the first kind.
The length $d(z,b(z'))$ is the length of the path connecting
$z$ and $z'$ with the topology according to $b\in \Gamma$
which is non-periodic for $z=z'$ in general.
Only in the special case in which $z=z'$ lies on the periodic orbit
belonging to $b$, the length of a periodic orbit is obtained.
It is the lengths of these periodic orbits
which are so peculiar degenerated in arithmetic systems.
However, since for the description of the wave functions according to
(\ref{Wave_Sum})
the distribution of the lengths $d(z,b(z'))$ is important,
the crucial question is, whether these lengths are,
for general $z$ and $z'$, differently distributed in arithmetic
and non-arithmetic systems.

\subsection{The distribution of $d(z,b(z))$}

To emphasize the role of $d(z,b(z))$ we specialize (\ref{Wave_Sum}) to a single
wave function by choosing as in \cite{Aurich91}
\begin{equation}
h(p) \; = \; \coth(\pi p) \, e^{-(p-p')^2/\varepsilon^2} \; - \;
\coth(\pi p) \, e^{-(p+p')^2/\varepsilon^2}
\hspace{10pt} , \hspace{10pt} \varepsilon > 0
\hspace{10pt} .
\end{equation}
Approximating the Mehler transform for large $p$ one gets the
semiclassical expression (eq.(37) in \cite{Aurich91})
\begin{equation}
\label{wfk_semicalssical_gauss}
\sum_{n=0}^\infty \; h(p_n) \; \left| \Psi_n(z) \right|^2 \; = \;
\frac{p\hspace{1pt}'\, \varepsilon}{\sqrt{4\pi}} \; + \;
\sum_{b\in\Gamma'} \frac{\sqrt{p\hspace{1pt}'} \, \varepsilon
\sin(p\hspace{1pt}' \tau_b + \frac \pi 4)}
{\sqrt{2\pi^2 \sinh \tau_b}} \; e^{-(\tau_b \varepsilon/2)^2}
\hspace{10pt} ,
\end{equation}
where $\tau_b := d(z,b(z))$, and
$\Gamma'$ is the Fuchsian group without the identity element
which contribution has to be considered separately leading to the
first term on the right side.
The sum over $\Gamma'$ is absolutely convergent for $\varepsilon > 0$.
Choosing $\varepsilon$ sufficiently small such that on the left-hand-side
only the summand belonging to $p'=p_n$ contributes,
allows the computation of a wave function solely from the Fuchsian group
$\Gamma$.
Therefore, the modulus of a wave function $\left| \Psi_n(z) \right|$
is determined solely by the lengths $\tau_b$.

Thus to address the question whether there are differences between arithmetic
and non-arithmetic systems with respect to their wave functions,
one can also investigate the properties of the lengths $\tau_b$
of arithmetic and non-arithmetic systems.
The simplest statistic is the nearest-neighbour length spacing
distribution $P(\Delta \ell,z)$.
Since the considered systems are invariant under time reversal,
all lengths occur twice in the sum over $\Gamma$ which is expressed by
the fact that $\tau_b = \tau_{b^{-1}}$ with $b\, b^{-1} = {\bf 1}$.
Thus in the statistic only one of the two lengths has to be included.
The next step is the unfolding of the length spectrum using the
asymptotic law \cite{Aurich91}
\begin{equation}
N(z,z',\tau) \; \sim \; \overline{N}(z,z',\tau) \; = \;
\frac{1}{4} e^\tau
\hspace{10pt} , \hspace{10pt} \tau \to \infty
\end{equation}
where $N(z,z',\tau)$ is the staircase function
\begin{equation}
N(z,z',\tau) \; := \; \#\{ b\, |\, b\in\Gamma, d(z,b(z')) \leq \tau \}
\hspace{10pt} .
\end{equation}
The unfolded lengths are denoted by
$\ell(z,z') = \overline{N}(z,z',\tau_b)$.
For the asymmetric and the regular octagon
all elements $b \in \Gamma$ are computed which can be represented as
a product of the generators of the group with up to 12 factors.
Both groups contain only hyperbolic elements.
(Elliptic and reflection elements occur later, when we consider
the group for the desymmetrized systems.)
It turns out that the length spacing distribution $P(\Delta \ell,z)$
is for almost all $z$ conform to a Poisson distribution
$P_{\hbox{\scriptsize Poisson}}(\Delta \ell) = e^{-\Delta \ell}$.
In the case of the asymmetric octagon one finds strong deviations
from the Poisson distribution at positions of the elliptic points
of the desymmetrized system,
which are marked in figure \ref{kreis_m31_g01} as dots.
In order to test the agreement with the Poisson distribution
we apply the Kolmogorov-Smirnov test to the distributions
in dependence of the position $z$.
The results are shown in figure \ref{Pic_Orbit_KS-test}.
Here the reverse of the significance level ${\cal P}$ is plotted,
i.\,e.\ the low regions are the domains with a Poisson distributed
length spacing.
It is clearly seen that in the case of the asymmetric octagon
only the elliptic points show strong non-Poisson behaviour
whereas the regular octagon displays also deviations along
symmetry lines due to the reflection elements
which tessellate the regular octagon in just the domains of the
triangular billiard.
However, inside the triangular billiard the lengths $\tau_b$ are
Poisson distributed within the same significance level
as in the asymmetric octagon excluding the elliptic points.

Recalling that the main difference between arithmetic and
non-arithmetic systems arises from the exponentially increasing
degenerations among the periodic orbits with respect to their lengths,
one now sees that such a behaviour does not occur with respect
to $\tau_b$ for almost all $z$.
Thus, no difference is expected between the wave functions of arithmetic
and non-arithmetic systems.
There might be, however, the possibility that the arithmetic structure
of the group leads to higher correlations among the $\tau_b$'s
which are independent of $z$ and simultaneously affect the sum
on the right-hand-side of eq.(\ref{wfk_semicalssical_gauss}),
but this seems to be unlikely.

\subsection{Corrections to $\overline{\sigma}_A$}

The other important application of the orbit theory is the derivation
of corrections to the leading term $\overline{\sigma}_A$ in terms
of the groups $\Gamma$.
Let us now consider the groups for the desymmetrized systems.
In the case of the asymmetric octagon the group is extended by
$S := \left( \begin{array}{rr} -i & 0 \\ 0 &i \end{array} \right)$
being a realization of the parity operation $z\to -z$
\cite{Aurich92}.
This group contains also elliptic elements.
The triangular billiard is described by the reflection group
which is generated by the three reflection elements which
realize the reflections along the three edges of the billiard.
This group contains in addition to the elliptic elements also
reflection elements.
(For more details, see e.\,g.\ \cite{Ninnemann95}.)
The formula eq.(\ref{Wave_Sum}) is also valid for these groups
if one takes for $\Gamma$ the above groups and takes the character $\chi(b)$
of the group elements $b\in\Gamma$ on the right-hand-side of (\ref{Wave_Sum})
into account.

To obtain the corrections to $\overline{\sigma}_A$ we choose in
eq.(\ref{Wave_Sum})
\begin{equation}
h(p) \; = \; \Theta( p_{\hbox{\scriptsize max}} - |p|\, )
\hspace{10pt} ,
\end{equation}
where $\Theta$ is the Heaviside step function.
This leads to
\begin{eqnarray}
\nonumber
\sum_{n=0}^N \; \left|\, \Psi_n(z)\, \right|^2 & = &
\frac{1}{2\pi} \, \int_0^{p_{\hbox{\scriptsize max}}} \; dp \;
p \tanh(\pi\,p) \\
\label{integral_cut}
& & \hspace{30pt} \; + \;
\frac{1}{2\pi} \; \sum_{b\in \Gamma'} \; \chi(b) \;
\int_0^{p_{\hbox{\scriptsize max}}} \; dp \;
p \tanh(\pi\,p) \, P_{-\frac 1 2 + ip}(\cosh(\tau_b)\,)
\hspace{10pt} ,
\end{eqnarray}
with $N$ such that $p_N \leq p_{\hbox{\scriptsize max}} < p_{N+1}$.
The left-hand-side is now rewritten as a Riemann-Stieltjes integral
\begin{equation}
\label{sum_cut}
\sum_{n=0}^N \; \left|\, \Psi_n(z)\, \right|^2 \; = \;
\int_0^{p_{\hbox{\scriptsize max}}} \; \left|\, \Psi_n(z)\, \right|^2 dN(p)
\; \simeq \;
\int_0^{p_{\hbox{\scriptsize max}}} \; \left|\, \Psi_n(z)\, \right|^2
d\overline{N}(p)
\hspace{10pt} ,
\end{equation}
where in the last step the approximation
$$
dN(p) \; \simeq \; d\overline{N}(p) \;  = \;
\left( \frac{\hbox{Area}({\cal M})}{2\pi} p \, + \,
\frac{{\cal L}}{4\pi} \,
\right) \; dp
$$
has been used together with Weyl's law including the circumference term.
The length ${\cal L}$ is defined as  ${\cal L} := {\cal L}^+ - {\cal L}^-$
whereby ${\cal L}^+$ and ${\cal L}^-$ are the lengths of the boundary
on which Neumann and Dirichlet boundary conditions are imposed,
respectively.
The comparison of the integrand of (\ref{integral_cut}) with (\ref{sum_cut})
yields
\begin{equation}
\label{Wave_single}
\left|\, \Psi_n(z)\, \right|^2 \; = \;
\frac{\tanh(\pi\,p_n)}{\hbox{Area}({\cal M}) \, + \,
\frac 1 {2p_n}{\cal L}} \;
\left\{ \; 1 \; + \; \sum_{b\in \Gamma'} \; \chi(b) \;
P_{-\frac 1 2 + ip_n}(\cosh(\tau_b)\,) \; \right\}
\hspace{10pt} .
\end{equation}
In contrast to (\ref{wfk_semicalssical_gauss}) this expression is at most
conditionally convergent.
Integrating equation (\ref{Wave_single}) over a domain ${\cal X}$ leads to an
orbit expression for $(A\Psi_n,\Psi_n)$ as defined in (\ref{DefScalar}).
The term corresponding to the identity element in eq.(\ref{Wave_single})
yields $\overline{\sigma}_A$.
The influence of the boundary, i.\,e.\ of the reflection elements,
and of the corners, i.\,e.\ of the elliptic elements, vanishes
in the semiclassical limit such that the properties of the $a_n$'s
are dominated by the contribution of the hyperbolic elements.
Thus one possible extrapolation to the semiclassical limit is provided
by introducing
\begin{equation}
\label{sigma_corr}
\overline{\sigma}_A^{\hbox{\scriptsize corr}} \; = \;
\left\{ \overline{\sigma}_A \; + \; \frac{1}{\hbox{Area}({\cal M})} \,
\sum_{b \in \Omega} \; \chi(b) \;
\int_{\cal X} \; d\mu(z) \; P_{-\frac 1 2 + ip_n}(\cosh(\tau_b)\,) \right\}
\; \left(\, 1 - \frac{{\cal L}}
{2\, p_n\, \hbox{Area}({\cal M})}\, \right)
\end{equation}
The definition of $\Omega \subset \Gamma'$ is arbitrarily and depends on the
effects one wishes to incorporate in $\overline{\sigma}_A^{\hbox{\scriptsize
corr}}$.
In the case of the two triangular billiards ${\cal A}$ and ${\cal B}$ only
the three reflection elements which realize the reflections at the boundary
of the billiards are included in our consideration.
It turns out that they cause the main contribution and that the contribution
of further reflection elements as well as of elliptic elements are negligible
for the domains considered here.
However, choosing domains which contain elliptic points leads to important
contributions
from just the corresponding elliptic elements.
In the case of the asymmetric octagon one has no reflection elements
such that the elliptic elements have to be considered.
The next class are the hyperbolic group elements whose inclusion in $\Omega$
is a matter of point of view.
Since the orbits spread uniformly over the phase space in ergodic systems
one suspects that they are responsible for the quantum ergodicity of the
wave functions and thus should not be included in (\ref{sigma_corr})
since it is just this effect one wants to study.

The effect of the reflection elements is well demonstrated by the quantity
\begin{equation}
\label{Def_kappa}
\kappa_n \; := \; \frac{\left(\, A \Psi_n, \Psi_n\right)}{\overline{\sigma}_A}
\end{equation}
and its average $\left<\kappa_n\right>$,
where $\left<\dots\right>$ denotes the averaging with the same
triangular shaped window function as above.
According to (\ref{Sarnak_Conjecture}) this tends towards one.
The result is shown in figure \ref{Pic_an_orbit} for the asymmetric octagon and
for
both billiards using the domains ${\cal X}$ shown in figure
\ref{kreis_m31_g01}.
As expected, the deviations from one are the larger the smaller $n$ is,
showing directly the influence of the boundary.
In the case of the asymmetric octagon the deviations are negligible for $n>
700$
showing only a minute influence of the elliptic elements.
The larger deviations of the two billiards are well explained by the
contribution
of the reflection elements.
Integrating (\ref{Wave_single}) over the domain ${\cal X}$ and taking only
the reflection elements into account provides the dashed curves shown in
figure \ref{Pic_an_orbit}.
Not only the mean behaviour of the deviations are explained
but also the long-range oscillations.
Subtracting this contribution from the $\kappa_n$ and averaging the result
gives the dotted curves in figure \ref{Pic_an_orbit}.
These fluctuate around one already for very small values of $n$.
The remaining fluctuations are due to the contributions of other group
elements not included in $\Omega \subset \Gamma'$.
The order of the amplitude of these fluctuations has to be compared with the
standard deviation $\Delta\kappa$ of the fluctuations of the $\kappa_n$'s
which are not averaged.
For billiard ${\cal A}$ and ${\cal B}$ one gets $\Delta\kappa=0.127$ and
$\Delta\kappa=0.106$, respectively,
which is orders of magnitudes larger than the range shown in
figure \ref{Pic_an_orbit}.
Computing $c_n^{\hbox{\scriptsize corr}} :=
| (A\Psi_n,\Psi_n)-\overline{\sigma}_A^{\hbox{\scriptsize corr}} | E^{1/4}$
and applying the fit (\ref{fit_erfc}) as in section \ref{behaviour_S_1}
gives for the $c_n^{\hbox{\scriptsize corr}}$'s for billiard
${\cal A}$ $\beta=0.5151\dots$
and $\delta=-0.0127\dots$ and for billiard ${\cal B}$ $\beta=0.4333\dots$
and $\delta=0.0018\dots$, respectively.
Thus the width of the distribution is unchanged but the shift $\delta$ of
the Gaussian is drastically reduced.
The significance levels turn out to be unaltered.

We have computed $S_1(E;A)$ with these corrected $a_n$'s but found
no significant improvement concerning the fit values of $\rho$.
It thus seems, that for some other reasons the two billiards approach slower
the suggested value $\frac 1 4$ than the asymmetric octagon.

\section{Summary and discussion}

In this paper we have investigated the behaviour of the wave functions
of three strongly chaotic (Anosov) systems.
The rate of quantum ergodicity is studied by the limit $E\to \infty$ of
$S_1(E;A)$ defined in (\ref{Def_Sk}) and (\ref{DefScalar}) and is found
to be consistent with the conjectured decline proportional to $E^{-1/4}$.
The average of the individual summands $a_n$, occurring in the sum in
(\ref{Def_Sk}),
is shown to decline also proportional to $E^{-1/4}$.
The $\tilde c_n := \tilde a_n E_n^{1/4}$, $\tilde a_n$ defined in
eq.(\ref{Def_an}),
are distributed according to a Gaussian with a non-zero mean for
finite energies.
Addressing the unique quantum ergodicity hypothesis we find no significant
exceptional behaviour in individual $\tilde c_n$'s, whereby we choose domains
far away from elliptic points and boundary lines of the considered systems.
These regions play a special role as it is demonstrated by the sum
(\ref{WFK_Sum})
of the modulus of the wave functions as seen in figure \ref{Pic_WFK_Sum}.
Sequences of domains ${\cal X}_i$ with increasing area are constructed
for which the computed $S_1(E;A)$ are compared with $f(E) = \alpha E^{-\rho}$.
The values of the fit parameter $\rho$ are consistent with the expected value
$\frac 1 4$.
The dependence of $\alpha$ on Area$({\cal X}_i)$ is linear for
Area$({\cal X}_i) \ll \hbox{Area}({\cal M})$ and declines thereafter towards
zero for Area$({\cal X}_i) \to \hbox{Area}({\cal M})$ after passing a
transition region.

The above analysis shows no differences between the arithmetic and the
non-arithmetic systems.
The orbit theory, which allows to express the wave functions in terms
of classical orbits, is applied to individual wave functions showing
that the modulus of the wave functions depends only on the hyperbolic distances
$\tau_b := d(z,b(z))$, see eq.(\ref{wfk_semicalssical_gauss}).
Since the peculiarities of the statistics of the quantal levels of
arithmetic systems are traced back to the exponential degeneration
of the lengths of the periodic orbits,
one has to pose the analogous question with respect to the wave functions.
In this case it is, however, the statistical properties of the lengths $\tau_b$
which determine possible peculiarities in arithmetic systems.
The nearest-neighbour length spacing $P(\Delta\ell,z)$ of the unfolded
$\tau_b$'s is Poisson distributed within the same significance
level in all three systems.
Only at the elliptic points of the asymmetric octagon and along the
boundaries of the billiards the distribution is non-Poisson,
as the Kolmogorov-Smirnov test presented in figure \ref{Pic_Orbit_KS-test}
shows.
Thus no peculiarities for the wave functions should be expected in arithmetic
systems
in contrast to the quantal levels.

The deviations of $(A\Psi_n,\Psi_n)$ from $\overline{\sigma}_A$ due to the
influences of the boundary conditions are computed in terms of the reflection
elements in the case of the two billiards.
The result presented in figure \ref{Pic_an_orbit} shows that the main
contribution
arises from the three reflection elements realizing the reflection at the
boundaries of the triangular billiards.

Therefore, our analysis gives strong numerical support for the quantum unique
ergodicity hypothe\-sis in the case of dynamical systems defined on
hyperbolic surfaces independent of arithmetic or non-arithmetic properties.


\vspace{1cm}

\leftline{\large \bf Acknowledgments}

We would like to thank the HLRZ at J\"ulich, the HLRS at Stuttgart
and the Rechenzentrum of the University Karlsruhe
for the access to their computers.
The figures \ref{Pic_WFK_Sum} and \ref{Pic_Orbit_KS-test} are computed using
{\it ``Blue Moon Rendering Tools''} written by L.\,Gritz and
{\it ``Geomview''} from the ``The Geometry Center'' of the University of
Minnesota.
Furthermore, we would like to thank F.\,Steiner, A.\,B\"acker and
R.\,Schubert for helpful discussions.


\setcounter{totalnumber}{20}
\setcounter{topnumber}{8}
\setcounter{bottomnumber}{8}

\newpage

\Large{\underline{Figure captions:}

\begin{figure}[hhh] 
\caption{\label{kreis_m31_g01}
On the left side the fundamental domain ${\cal M}$ of the asymmetric octagon is
shown on the Poincar\'e disc ${\cal D}$.
The involution symmetry is clearly visible.
The chosen operator $A$ acts non-vanishingly on the grey disc ${\cal X}$.
The dots denote the position of the elliptic points on this surface.
On the right side the triangular billiard is presented with the two
boundary conditions chosen here.
}
\end{figure}

\begin{figure}[hhh] 
\caption{ \label{an_sarnak_conjecture}
The average $\left< a_n \right>$ over 100 neighbouring values
for the domains ${\cal X}$ shown in figure \ref{kreis_m31_g01}.
The corresponding fits $f(E) = c_f E^{-1/4}$ are shown as dashed curves.
}
\end{figure}

\begin{figure}[hhh] 
\caption{ \label{cn_m31_g01}
The normalized pseudo-random numbers $\tilde c_n$ are displayed for the
asymmetric octagon.
}
\end{figure}

\begin{figure}[hhh] 
\caption{ \label{Pic_WFK_Sum}
$W(z,E)$ is shown for the asymmetric octagon, for billiard ${\cal A}$
and for billiard ${\cal B}$.
The triangular billiards are rotated such that one looks on the Dirichlet
boundary.
For the asymmetric octagon we have $E=6\,000$ and for the two billiards
$E=200\,000$.
}
\end{figure}

\begin{figure}[hhh] 
\caption{ \label{Distr_cn_m31_g01}
The cumulative distribution $I(\tilde c\,)$ of the normalized numbers
$\tilde c_n$ is shown in comparison with a cumulative Gaussian (dashed curve)
for the same operator $A$ as in figure \protect\ref{cn_m31_g01}.
}
\end{figure}

\begin{figure}[hhh] 
\caption{ \label{S1_Oktagon_G01}
$S_1(E;A)$ is shown for the asymmetric octagon as a full curve.
The dashed curve is obtained by evaluating the sum
with the mean values $\overline{a}_n$ depending only on $\overline{c}$.
However, the first 50 correct terms $a_n$ have been included,
see eq.(\protect\ref{S1_prakt}).
The dotted curves correspond to one standard deviation from the mean
behaviour thus demonstrating the decline with the power $\frac 1 4$.
}
\end{figure}

\begin{figure}[hhh] 
\caption{ \label{Pic_circle_series}
The chosen sequences of domains ${\cal X}_i$ are shown for the considered
systems.
}
\end{figure}

\begin{figure}[hhh] 
\caption{ \label{Pic_S1_series}
$S_1(E;A)$ and the fit (\protect\ref{Fit}) are shown for the
asymmetric octagon in a), for billiard $\protect{\cal A}$ in b)
and for billiard $\protect{\cal B}$ in c).
The 10 different curves belong to the 10 domains shown in
figure \protect\ref{Pic_circle_series}.
}
\end{figure}

\begin{figure}[hhh] 
\caption{ \label{Pic_alpha}
The fit parameter $\alpha$ is plotted for the three systems in dependence
on the area of the domain ${\cal X}_i$.
The full curve corresponds to the asymmetric octagon, the dotted curve
to billiard ${\cal A}$ and the dashed curve to billiard ${\cal B}$.
}
\end{figure}

\begin{figure}[hhh] 
\caption{ \label{Pic_Orbit_KS-test}
The reverse of the significance level ${\cal P}$ that the unfolded
length spacing of $d(z,b(z))$ is distributed according to a
Poisson distribution, is shown.
}
\end{figure}

\begin{figure}[hhh] 
\caption{ \label{Pic_an_orbit}
The quantity $\left< \kappa_n \right>$ is shown for the asymmetric octagon
(dashed-dotted curve) and for both billiards (full curves) for the domains
${\cal X}$ shown in figure \ref{kreis_m31_g01}.
The lowest curve belongs to billiard ${\cal B}$, and the curve belonging
to billiard ${\cal A}$ lies above it.
The contribution of the identity and the three most important reflection
elements according to (\ref{Wave_single}) is displayed as dashed curves.
Subtracting this contribution from $\kappa_n$ and averaging the result yields
the dotted curves.
}
\end{figure}

\end{document}